\documentclass[aps,prl,twocolumn,superscriptaddress]{revtex4}
\usepackage{graphicx}
\usepackage{color}

\begin{document}
\title{Ultrafast Dynamics of Massive Dirac Fermions in Bilayer Graphene}

\author{S\o ren Ulstrup}
\affiliation{Department of Physics and Astronomy, Interdisciplinary Nanoscience Center (iNANO), Aarhus University, 8000 Aarhus C, Denmark}
\author{Jens Christian Johannsen}
\affiliation{Institute of Condensed Matter Physics, \'Ecole Polytechnique F\'ed\'erale de Lausanne (EPFL), 1015 Lausanne, Switzerland}
\author{Federico Cilento}
\affiliation{Sincrotrone Trieste, 34149 Trieste, Italy}
\author{Jill A. Miwa}
\affiliation{Department of Physics and Astronomy, Interdisciplinary Nanoscience Center (iNANO), Aarhus University, 8000 Aarhus C, Denmark}
\author{Alberto Crepaldi}
\affiliation{Sincrotrone Trieste, 34149 Trieste, Italy}
\author{Michele Zacchigna}
\affiliation{IOM-CNR Laboratorio TASC, Area Science Park, 34012 Trieste, Italy}
\author{Cephise Cacho}
\author{Richard Chapman}
\author{Emma Springate}
\affiliation{Central Laser Facility, STFC Rutherford Appleton Laboratory, Didcot OX11 0QX, United Kingdom}
\author{Samir Mammadov}
\author{Felix Fromm}
\author{Christian Raidel}
\author{Thomas Seyller}
\affiliation{Institut f\"ur Physik, Technische Universit\"at Chemnitz, 09126 Chemnitz, Germany}
\author{Fulvio Parmigiani}
\affiliation{Sincrotrone Trieste, 34149 Trieste, Italy}
\affiliation{Department of Physics, University of Trieste, 34127 Trieste, Italy}
\author{Marco Grioni}
\affiliation{Institute of Condensed Matter Physics, \'Ecole Polytechnique F\'ed\'erale de Lausanne (EPFL), 1015 Lausanne, Switzerland}
\author{Phil D. C. King}
\affiliation{SUPA, School of Physics and Astronomy, University of St. Andrews, St. Andrews, Fife KY16 9SS, United Kingdom}
\author{Philip Hofmann}
\affiliation{Department of Physics and Astronomy, Interdisciplinary Nanoscience Center (iNANO), Aarhus University, 8000 Aarhus C, Denmark}
\email[]{philip@phys.au.dk}

\date{\today}
\begin{abstract}
Bilayer graphene is a highly promising material for electronic and optoelectronic applications since it is supporting massive Dirac fermions with a tuneable band gap. However, no consistent picture of the gap's effect on the optical and transport behavior has emerged so far, and it has been proposed that the insulating nature of the gap could be compromised by unavoidable structural defects, by topological in-gap states, or that the electronic structure could be altogether changed by many-body effects. Here we  directly follow the excited carriers in bilayer graphene on a femtosecond time scale, using ultrafast time- and angle-resolved photoemission. We find a behavior consistent with a single-particle band gap. Compared to monolayer graphene, the existence of this band gap leads to an increased carrier lifetime in the minimum of the lowest conduction band. This is in sharp contrast to the second sub-state of the conduction band, in which the excited electrons decay through fast, phonon-assisted inter-band transitions.
\end{abstract}

\maketitle
The lack of a band gap is the most important obstacle to using graphene in electronic devices but this can be elegantly solved in bilayer graphene (BLG) when an asymmetry between the layers is induced by a transverse electric field \cite{mccann:2006,McCann:2006b,ohta:2006}. The promising properties of the thereby induced massive Dirac particles have been intensively explored for the development of semiconducting devices with tuneable band gaps \cite{castro:2007,oostinga:2008,zhang:2009} and for efficient photodetectors extending to the THz regime \cite{xia:2009,wright:2009,bonaccorso:2010,yan:2012}. However, the quantitative transport properties of BLG are inconsistent with the simple scenario of a small band gap semiconductor \cite{oostinga:2008} and different hypotheses for this have been given, including broken-symmetry ground states \cite{Weitz:2010} or topological edge state effects \cite{Li:2011}. Recently, a study of BLG by angle-resolved photoemission spectroscopy (ARPES) has revealed the presence of electronic states throughout the gap, arising from intrinsic AA-stacked domains \cite{kim:2013}, something that might be expected to short-circuit the gap of BLG.

While static ARPES results provide crucial information about the spectral function of BLG, time-resolved spectroscopy is needed to complement the transport studies. Time- and angle-resolved photoemission (TR-ARPES) experiments near the Dirac cone have only recently become technically feasible \cite{johannsen:2013,gierz:2013} due to the high photon energies needed to reach the $\bar{K}$ point in the Brillouin zone. Here we report TR-ARPES measurements carried out using a Ti:sapphire amplified laser system with a repetition rate of 1 kHz. This provided ultrafast infrared pulses with a wavelength of 785 nm, a full width at half-maximum (FWHM) duration of 30 fs, and an energy per pulse of 12 mJ. A part of the laser energy was applied for high harmonic generation
of extreme ultraviolet pulses in a pulsed jet of argon gas. A time-preserving monochromator was used to single-out the 13th harmonic with a photon energy of 21~eV. 
The remaining laser energy was utilized to drive an optical parametric amplifier (HE-Topas), which can provide tuneable pump pulses from the UV to the mid-infrared. We used the second harmonic of the signal to produce 30 fs pulses at 1.55 eV. Both beams were polarized perpendicular to the scattering plane ($s$-polarized) The data were acquired by sweeping 45 time delay points. Such a cycle was repeated between 700 and 1000 times in order to obtain satisfactory statistics in the data, amounting to total acquisition times between 8 and 14~h per dataset. The experiments were carried out at the Artemis facility at the Central Laser Facility, Rutherford Appleton Laboratory\cite{Frassetto:2011}.

\begin{figure*}
\includegraphics[width=.95\textwidth]{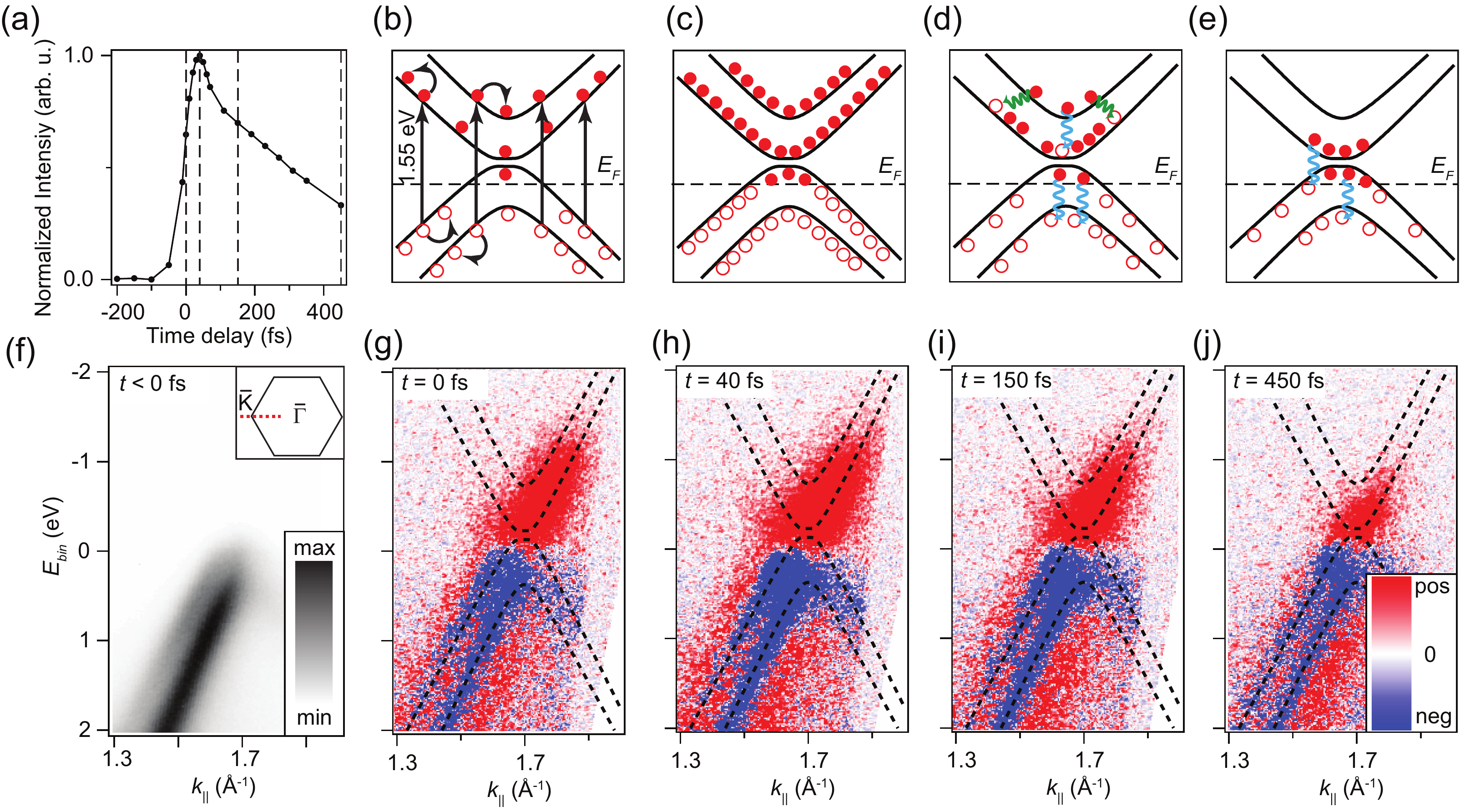}\\
\caption{(Color online) Ultrafast ARPES measurements of bilayer graphene: (a) Normalized photoemission intensity integrated above the Fermi level. Vertical dashed lines mark the times of the dispersion snapshots in (g)-(j). (b)-(e) Diagrams of ultrafast processes and relaxation dynamics involving optical pumping (straight arrows), electron scattering (curled arrows), optical (blue wiggled arrows) and acoustic (green wiggled arrows) phonon scattering. Filled (open) circles signify electrons (holes). (f) Spectrum before the arrival of the pump pulse with the inset showing the acquisition direction for all data ($\bar{\Gamma}-\bar{K}$). (g)-(j) ARPES difference spectra obtained by subtracting the spectrum in (f) from the spectrum taken at the given time. A pump laser fluence of $F=1.0$~mJ/cm$^2$ was used. The dashed parabolic bands are the result of a tight binding calculation, and have been added as a guide to the eye.}
  \label{fig:1}
\end{figure*}

Samples of high quality, so-called quasi-free-standing mono- and bilayer graphene were prepared ex situ
by thermal decomposition of SiC substrates followed by hydrogen intercalation to
decouple the graphene layers from the substrate. The method is described in detail in Ref. \cite{Speck:2010}.
Both samples are hole doped with carrier concentration on the order of $5\times10^{12}$~cm$^{-2}$, which places the Dirac point 240~meV above the Fermi level in the monolayer sample, while in the bilayer the center of the band gap is 180~meV above the Fermi level. Both samples were transferred through air into the TR-ARPES ultrahigh vacuum end station, where they were cleaned by annealing to 550~K in order to remove adsorbed impurities. The samples were held at 90~K throughout the experiment.
 
Fig. \ref{fig:1}(a) presents the integrated intensity of photo-induced excited states above the Fermi level during and immediately after the pumping phase. The pump signal induces direct transitions from the two parabolic valence bands to their conduction band counterparts. This pumping phase is characterized by a fast rising edge (see Fig. \ref{fig:1}(a)) that establishes the time resolution of our experiment to be 40~fs. The 
excited electrons quickly thermalize, consistent with rapid Auger-like processes \cite{Rana:2007,brida:2013,johannsen:2013} as depicted in Fig. \ref{fig:1}(b). This leaves a dense transient population of hot electrons in both conduction bands as sketched in Fig. \ref{fig:1}(c). The electrons then quickly cascade down to the region around the band gap, thereby depleting the conduction bands by emission of phonons (Fig. \ref{fig:1}(d)). This process continues (Fig. \ref{fig:1}(e)) until all of the excited electrons have recombined with the remaining holes.

Snapshots of these events are presented in Figs. \ref{fig:1}(g)-(j), displaying the ARPES difference spectra between the equilibrium signal before the arrival of the pump (Fig. \ref{fig:1}(f)) and the signal at the given time delay.  The bare band dispersion lines obtained by a tight binding calculation (see Ref. \cite{mccann:2006} and Supplementary Material) are plotted on top of the difference spectra to match the intensity with the band structure. The two bilayer bands are clearly discerned in the raw data as well as in the valence band part of the difference spectra. Note that we do not observe discrete spikes in the intensity distribution at the energies corresponding to the direct transition, not even on the midpoint of the rising edge at $t=0$~fs (see Fig. \ref{fig:1}(g)), but rather an even intensity distribution along the bands. This implies that the initial thermalization of the hot electrons by electron-electron scattering occurs on timescales less than 40~fs, too fast to be resolved here. Once the peak distribution is reached at $t=40$~fs (see Fig. \ref{fig:1}(h)), the phonon scattering processes lead to a fast decrease of intensity as observed in Figs. \ref{fig:1}(i)-(j).

\begin{figure}
\includegraphics[width=.49\textwidth]{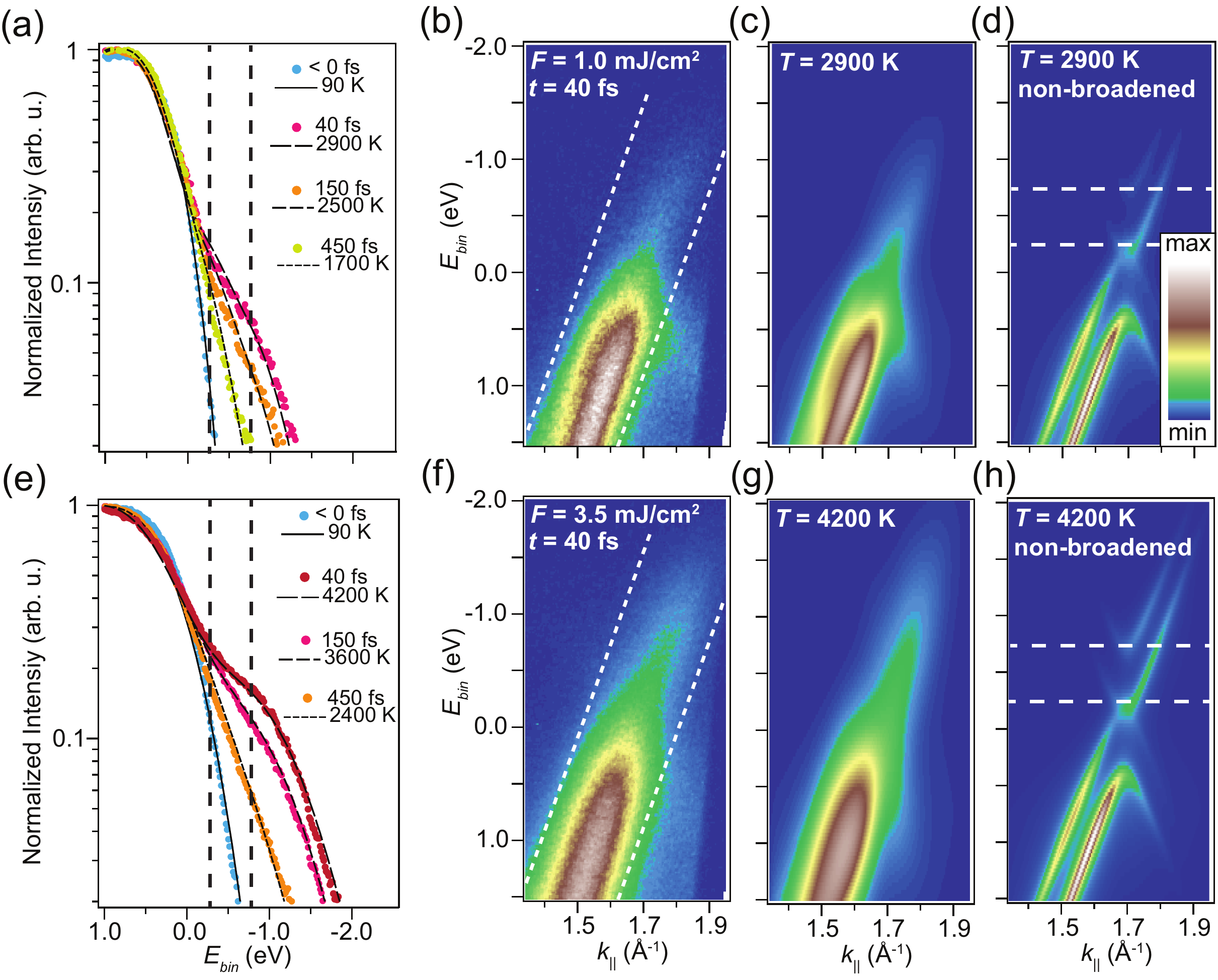}\\
\caption{(Color online) Disentangling the dynamics in the bilayer bands: (a), (e) Photoemission intensity integrated between the dashed lines in (b) and (f) as a function of binding energy. Markers correspond to data points at the given time delay, and lines are integrated intensity from simulated photoemission data with the given temperatures. (b), (f) Experimental photoemssion data at the peak excitation signal at time $t = 40$~fs. (c), (g) Simulated data at the given temperature taking the experimental resolution into account. (d), (h) Simulated data without resolution broadening. The vertical dashed lines in (a) and (e) and the horizontal dashed lines in (d) and (h) mark the onset energy of the two conduction bands. The data in (a)-(d) correspond to a fluence of 1.0~mJ/cm$^2$ while the data in (e)-(h) correspond to 3.5~mJ/cm$^2$.}
  \label{fig:2}
\end{figure}

Within the considered time window, the described behavior is qualitatively identical  for the laser pump fluence of $F=$1.0~mJ/cm$^2$ used for acquiring the data in Fig. \ref{fig:1} and for another data set with $F=3.5$~mJ/cm$^2$ (see Supplementary Material for the high fluence data.). However, with the higher fluence a larger density of photoexcited electrons is created, which allows us to study a larger population of hot electrons in the two conduction bands. A quantitative study of this is introduced in Fig. \ref{fig:2}. Here we compare the time-dependent experimental data with simulations where we populate the conduction bands thermally by multiplying a model spectral function with a single high temperature Fermi-Dirac (FD) function (see Supplementary Material for details on the simulations).

We extract the combined statistical distribution of the carriers in the four bands at a given time delay by integrating the photoemission intensity in between the dashed lines in Figs. \ref{fig:2}(b) and \ref{fig:2}(f) \cite{Ulstrup:2014}. Interestingly, the integrated distribution curves (IDCs) of the transient spectra presented in Figs. \ref{fig:2}(a) and \ref{fig:2}(e) do not  appear to be merely described by a featureless hot electron FD tail, but a clear shoulder can be discerned in the raw data. The extent of this feature is a function of time and of fluence. As it is not possible to extract the temperature from the data by a fit to a simple FD function, we do so by matching the observed photoemission intensity to a simulation, consisting of the calculated spectral function that is multiplied by a hot FD function and broadened by the experimental resolution. Note that the intensity variation between the bands and within the bands originate from the photoemission matrix elements that are accounted for in the simulated spectral functions. These matrix elements are assumed to be energy- and band-dependent but independent of the fluence or time, as described further in the Supplementary Material. We find that the electrons are heated from the initial temperature of 90~K, which is the sample temperature during the experiment, to reach temperatures of 2900~K and 4200~K for low and high fluence, respectively. 

A remarkable agreement between experimental and simulated data is obtained at both low and high fluence. It is a central result that the complex shape of the IDCs from the peak signal at $t = 40$~fs and onwards in time can be described by this simple simulation. The observed shoulder in the IDCs always appears at the same energy which coincides with the onset of the inner conduction band, as seen by comparing the IDCs to the non-broadened BLG spectral function via the dashed lines in Fig. \ref{fig:2}. The shoulder can therefore be ascribed to the accumulation of electrons in the upper conduction band, confirming the consistency of the calculated spectral function with the data. The fact that all the IDCs can be fitted with this simple model also suggests that the excited carriers in both conduction bands are in a quasi-equilibrium state within our time resolution. 

\begin{figure*}
\includegraphics[width=.8\textwidth]{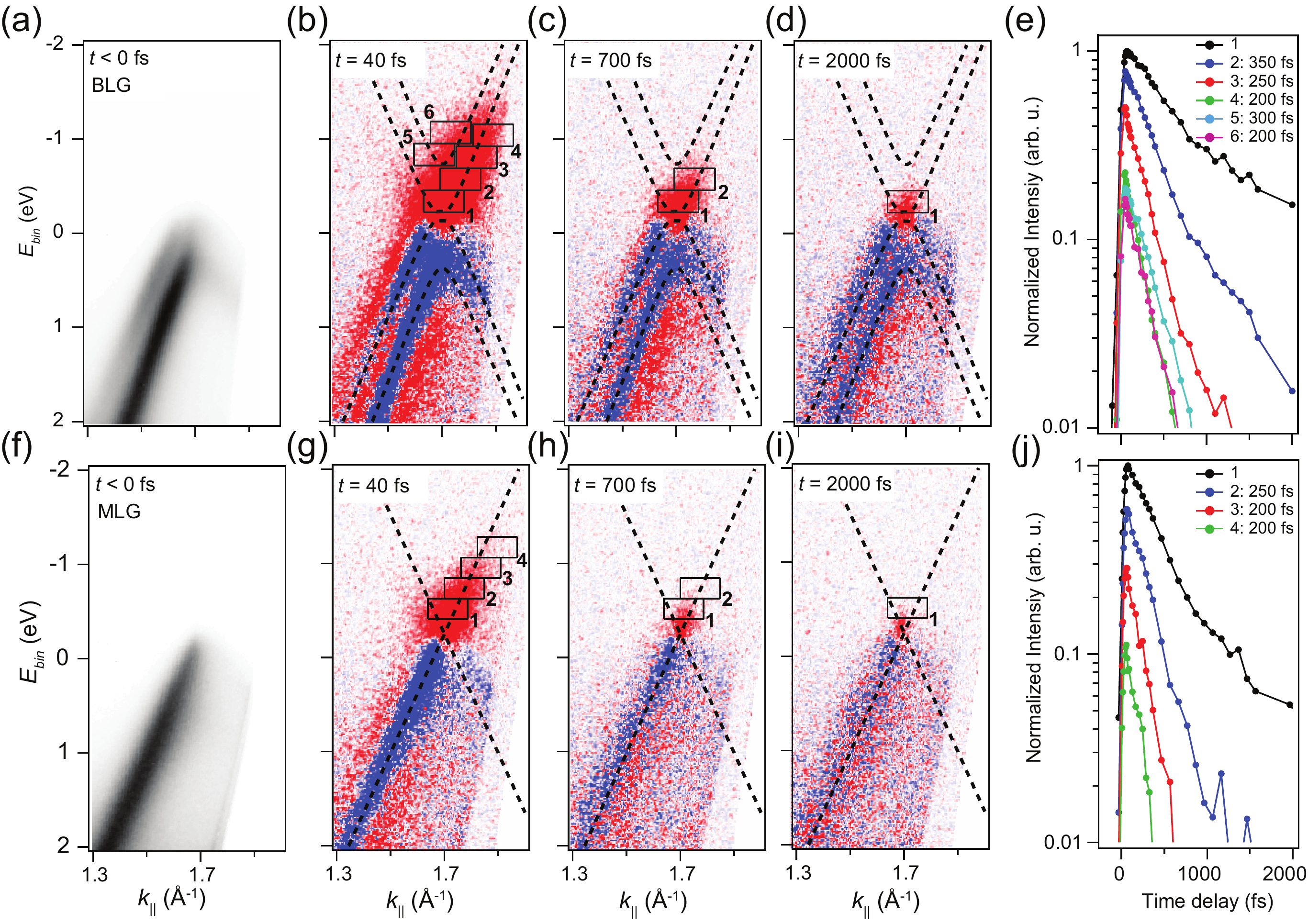}\\
\caption{(Color online) Comparing the relaxation dynamics of massive and massless Dirac Fermions: (a) Spectrum before the arrival of the pump pulse, and (b)-(d) difference spectra at selected time delays for BLG. (f)-(i) Corresponding data for MLG. Dispersion lines from tight binding models have been added as guide to the eye. (e), (j) Normalized photoemission intensity binned over the boxes in (b)-(d) for (e) BLG and (g)-(i) for (j) MLG. The time traces have been labelled with a number according to the box numbers, and the given time constants are results of single exponential fits to the corresponding trace. The error bar on these fits amounts to $\pm 20$~fs. The fluence is 0.6~mJ/cm$^2$ for all data presented here.}
  \label{fig:3}
\end{figure*}

The model presented in Fig. \ref{fig:2} integrates over all the sub-bands in the system and can thus neither be used to study possible differences in the sub-bands, nor describe lifetime effects that do not follow a FD distribution. 
A more detailed analysis and a direct comparison to the situation in MLG is therefore shown in Fig. \ref{fig:3}. For both materials, the Dirac point imposes a bottleneck for relaxation processes, but the key-consequence of a true band gap in BLG would be to impose a serious constriction for the decay of hot electrons, e.g. by scattering with low-energy acoustic phonons.     
To avoid any complications from high fluence effects, as described in the supplementary material, we consider both MLG and BLG at a low fluence of $0.6$~mJ/cm$^2$. The equilibrium spectra for BLG and MLG at this fluence are shown in Figs. \ref{fig:3}(a) and 3(f), and the difference spectra from $t=40$~fs  to long time delays of $t=2000$~fs are shown in Figs. \ref{fig:3}(b)-(d) and 3(g)-(i), respectively. Again we show the tight binding bands as a guide to the eye.  

The timescales of the relaxation processes are directly extracted by analysing the intensity binned in small boxes following the dispersion, as shown in the difference spectra in Fig. \ref{fig:3}. We emphasize that this analysis, contrary to the simple model in Fig. 2, provides access to the spectrally resolved dynamics at specified binding energies \cite{Crepaldi:2012}, capturing dynamics that would otherwise be integrated out using the model. The time-dependent intensity for all boxes is shown in Figs. \ref{fig:3}(e) and 3(j) and we find that it can be described by a simple exponential decay, apart from box 1, the box of the conduction band minimum in both samples (Note that the location of box 1 has been defined such that it is at equal distance from the center of the gap and the Dirac point in BLG and MLG, respectively). We shall discuss the intensity in this spectral region later. 

The electron population in the upper conduction band in BLG is tracked by boxes 5 and 6 in Fig. \ref{fig:3}(e) and we find a fast decay with a time constant of 300 and 200~fs, respectively. For box 6, this is expected because the electrons can cascade down to the minimum of the upper conduction band. For box 5, the fast decay can only be explained by phonon-induced inter-band scattering to the lower conduction band. A similarly fast decay as for boxes 5 and 6 is found for corresponding boxes in the lower conduction band (boxes 3-4), suggesting that the inter-band scattering is sufficiently effective to  avoid any electron accumulation near the upper conduction band edge. Indeed, this is also responsible for the fact that the simple FD fit works in Fig. \ref{fig:2}. The decay times for the corresponding energies in MLG are also similar. 

\begin{figure}
\includegraphics[width=.49\textwidth]{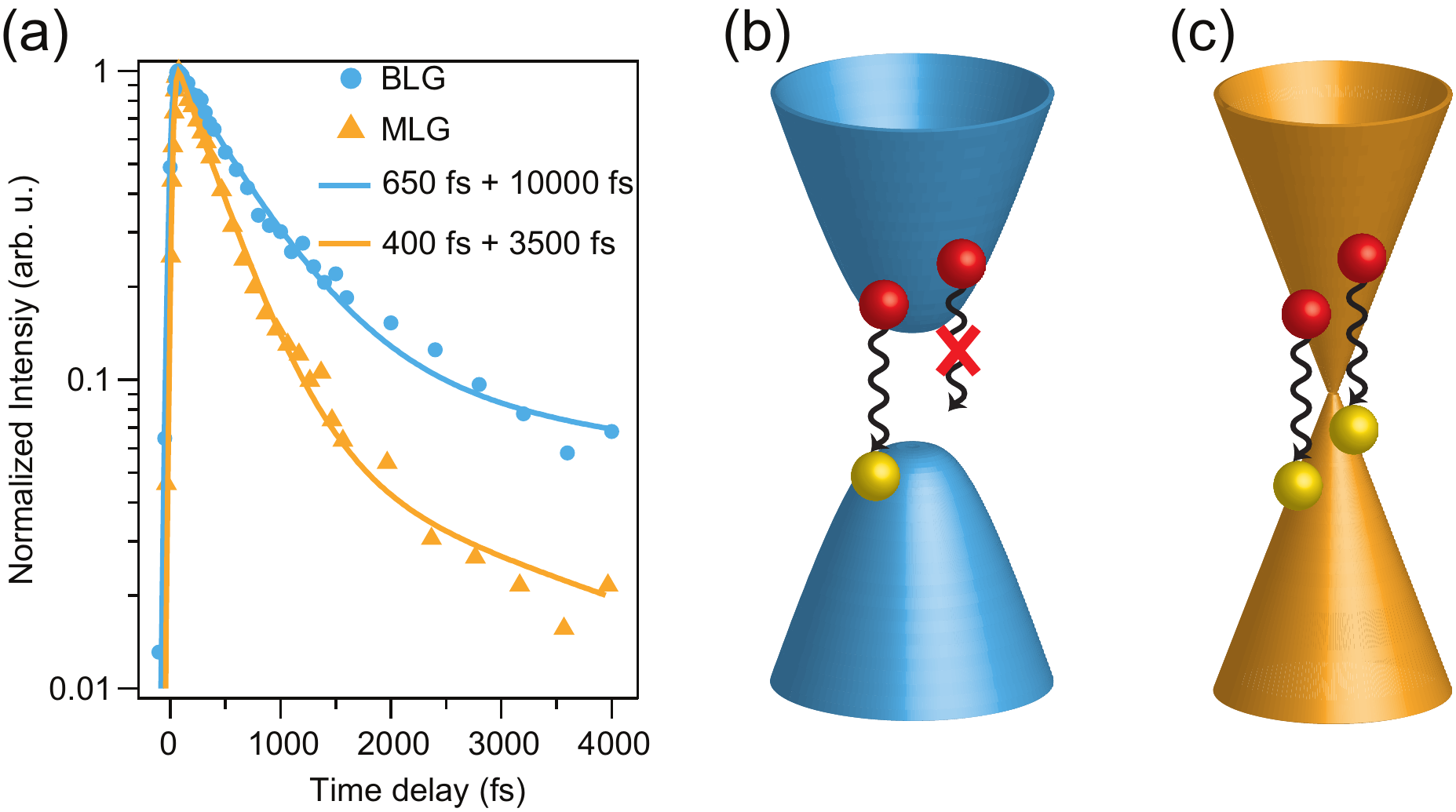}\\
\caption{(Color online) Gap dependent relaxation times: (a) Long time traces of the integrated intensity in box 1 in Fig. 3 for both BLG and MLG. Lines correspond to double-exponential function fits with the given time constants. (b), (c) Sketch of the relaxation dynamics involving hot electrons (red spheres), holes (yellow spheres) and phonon emission (wiggled arrows) around the Dirac point for (b) massive Dirac Fermions and (c) massless Dirac Fermions.}
  \label{fig:4}
\end{figure}

The situation is more complicated for box 1 in both materials where the intensity decay cannot be fit to a single exponential. The time-dependent intensity of box 1 is shown in Fig. \ref{fig:4} up to a measured delay time of 4~ps.  For this spectral region, the very bottom of the conduction band, pronounced differences for BLG and MLG are found. Both data sets can be well-described by a double-exponential decay where an initial fast decay with a decay time of 650~fs (400~fs) for BLG (MLG) is followed by a slow decay with 10,000~fs (3,500~fs). Indeed, the slow decay for BLG is so slow that a precise determination of the decay constant is not possible within our experimental time window and the value of 10,000~fs merely sets a lower limit for the long decay time. The pronounced difference between the two materials can actually already be seen in the raw data in Figs. \ref{fig:3}(d) and \ref{fig:3}(i). In BLG there is still intensity from hot electrons in box 1 while in MLG the whole spectrum is fully relaxed. 

The true difference in the hot electron carrier dynamics thus occurs  mainly at the Dirac point - an experimental verification that is only possible using this direct momentum-resolved technique.
This result is consistent with the naive expectation of a substantially longer decay time in the presence of a gap. Without a gap, hot electrons can cascade down continuously via acoustic phonon scattering whereas the presence of the gap requires relaxations by optical phonons, giving rise to a bottleneck for the decay. The efficiency of a relaxation across the gap might be further restricted by the role of the pseudospin in BLG \cite{mccann:2006}.   An intuitive picture of these differences between BLG and MLG is depicted in Figs. \ref{fig:4}(b) and (c). 

In view of the disputed character of the electronic structure of BLG and the role of possible in-gap states, we should also point out that the spectral function fits and the data analysis in Fig. \ref{fig:3} are perfectly consistent with the expected single-particle tight-binding band structure. The possibility of in-gap states still exists, especially that of gap-crossing states in AA-stacked domains, since these were identified for the same kind of BLG samples used here \cite{kim:2013}. In order to obtain a quantitative estimate of the impact these defects have on the lifetime of the carriers, a calculation of the carrier lifetime for the perfect crystalline system would be required. Nevertheless, the results shown here reveal that, even in the presence of such gap states, the band gap leads to a markedly enhanced lifetime over MLG.

We gratefully acknowledge financial support from the VILLUM foundation, The Danish Council for Independent Research / Technology and Production Sciences, the Lundbeck Foundation, the Swiss National Science Foundation (NSF), EPSRC, The Royal Society  and the Italian Ministry of University and Research (Grants No. FIRBRBAP045JF2 and No. FIRB-RBAP06AWK3). Access to the Artemis Facility was funded by STFC. Work in Erlangen and Chemnitz was supported by the European Union through the project ConceptGraphene, and by the German Research Foundation in the framework of the SPP 1459 Graphene.

\end{document}